\documentclass{article}
\usepackage{arxiv}
\pdfoutput=1

\usepackage[T1]{fontenc}    
\usepackage{hyperref}       
\usepackage{url}            
\usepackage{booktabs}       
\usepackage{amsfonts}       
\usepackage{nicefrac}       
\usepackage{microtype}      
\usepackage{lipsum}

\usepackage{expl3}
\usepackage{environ}
\ExplSyntaxOn
\seq_new:N \g_appendices_seq
\NewEnviron{Appendix}{\seq_gput_right:No \g_appendices_seq \BODY}
\newcommand\AddAppendices{
  \appendix
  \seq_map_inline:Nn \g_appendices_seq {##1}
}
\ExplSyntaxOff

\usepackage{etoolbox}
\AtEndDocument{\AddAppendices}

\usepackage{booktabs} 

\usepackage{standalone}

\usepackage{pgf}
\usepackage{tikz}
\usetikzlibrary{automata,shapes.multipart} %
\usetikzlibrary{arrows,petri}
\usepackage[latin1]{inputenc}
\usepackage{algorithm}
\usepackage{epsfig}
\usepackage{subfigure}
\usepackage{multirow}
\usepackage{graphicx}
\usepackage{color}
\usepackage{amsmath,amssymb,dsfont}
\usepackage{array}
\usepackage{multirow}

\usepackage{array}
\newcolumntype{L}[1]{>{\raggedright\let\newline\\\arraybackslash\hspace{0pt}}m{#1}}
\newcolumntype{C}[1]{>{\centering\let\newline\\\arraybackslash\hspace{0pt}}m{#1}}
\newcolumntype{R}[1]{>{\raggedleft\let\newline\\\arraybackslash\hspace{0pt}}m{#1}}

\colorlet{orange}{red!70!yellow}
\colorlet{vert}{green!70!blue}
\colorlet{mauve}{blue!70!red}
\colorlet{brouge}{red!70!blue}
\colorlet{frouge}{red!40!black}

\usepackage{url}
\urldef{\mailsa}\path|{alfred.hofmann, ursula.barth, ingrid.haas, frank.holzwarth,|
\urldef{\mailsb}\path|anna.kramer, leonie.kunz, christine.reiss, nicole.sator,|
\urldef{\mailsc}\path|erika.siebert-cole, peter.strasser, lncs}@springer.com|

\newcommand{\cosmos}{\mbox{\textup{C}\scalebox{0.75}{{\textsc{OSMOS}}}}}

\newcommand{\ignore}[1]{}

\title{Performance modelling of  access control mechanisms for  local  and vehicular wireless networks}

\author{
  Paolo Ballarini\thanks{Corresponding author.} \\
  Lab. MICS\\
  University Paris Saclay\\
  Gif-sur-Yvette, France. \\
  \texttt{paolo.ballarini@centralesupelec.fr} \\
   \And
 Beno\^it Barbot \\
  Lab. LACL\\
  University Paris Est Cr\'eteil\\
  Cer\'eteil, France\\
  \texttt{benoit.barbot@u-pec.fr} \\
   \AND
  Nicolar Vasselin \\
  CentraleSup\'elec \\
  Gif-sur-Yvette, France.  \\
   \texttt{nicolas.vasselin@student.ecp.fr} \\
}

\begin{document}
\maketitle

\begin{abstract}
Carrier sense multiple access collision avoidance (CSMA/CA)  is the basic scheme  upon which access to the shared medium is regulated in many wireless networks. With CSMA/CA a station willing  to start a transmission  has first to find  the channel free for a given duration otherwise it will go into \emph{backoff}, i.e. refraining for transmitting for a randomly chosen delay. Performance analysis of a wireless network employing  CSMA/CA regulation is not an easy task: except  for simple network configuration analytical solution of key performance indicators (KPI) cannot be obtained hence one has to resort to formal modelling tools. In this paper we present  a performance modelling study targeting different kind of CSMA/CA based wireless networks, namely:  the IEEE 802.11  Wireless Local Area Networks (WLANs) and   the 802.11p    Vehicular Ad Hoc Networks (VANETs), which extends 802.11 with priorities over packets. 
The modelling framework we introduce  allows for considering: i)  an arbitrarily large number of stations, ii) different traffic  conditions (saturated/non-saturated), iii) different hypothesis concerning the shared channel (ideal/non-ideal). We apply statistical model checking to assess KPIs of different network configurations.
\end{abstract}

\date{}

\section{Introduction}
\label{sec:introduction}

Communication protocols regulate the behaviour of 
communicating nodes within a concurrent environment. 
The Open System Interconnection (OSI) model \cite{dazi} defines a
layered architecture for network protocols. The {\it Medium Access
Control} (MAC) layer, part of the  data-link layer, determines which node is allowed to access the underlying physical-layer (i.e. the medium) 
at any given moment in time. A MAC scheme is mainly concerned with reducing the possibility of \emph{collisions} (i.e. simultaneous transmissions over a shared channel) from taking place. The basic mechanism used for reducing the likelihood of collisions, usually referred to as Carrier Sense Multiple Access (CSMA),  is  that,  before starting a transmission, any node should sense the medium clear for a given period.

\noindent
{\bf IEEE 802.11 WLAN}. 
Wireless local area networks (WLANs) are
wireless networks for which either the communication is managed by a
centralised Access Point (AP) or, in the case of  
{\it ad-hoc}, nodes communicate in a 
peer-to-peer fashion through a distributed coordination function. 
The IEEE 802.11 \cite{ieee80211_97} is a family of standards which specifies a number of MAC schemes and 
the {\it Physical} (PHY) layer for WLANs. 
The primary MAC scheme of the standard is called {\it Distributed Coordination 
Function} (DCF). It describes a de-centralised mechanism which allows network 
stations to coordinate for the use of a (shared) medium in an attempt to 
avoid collision. 
The DCF is a variant of  
the CSMA/CA MAC scheme developed for
collision avoidance over a shared medium using a randomised backoff procedure. 
Two variants of the DCF have been defined in the standard: 
the {\it Basic Access} (BA), which uses a single \emph{acknowledgement}  to confirm the successful reception of a data packet, and the {\it Request-To-Send/Clear-To-Send} (RTS/CTS), which employs a \emph{double-handshaking} scheme so to reduce costly collisions on large data packets. 

\noindent
{\bf IEEE 802.11p VANET}. 
In  the realm of Intelligent Transportation Systems (ITS) the main concern is one of improving the effectiveness as well as the safety of future transportation systems. ITS entail  \emph{hybrid}  communication scenarios where both   Inter-Vehicle-Communication (IVC), based on  \emph{ad-hoc} connections between moving vehicles, and  Roadside-Vehicle-Communication (RVC), concerned with the exchanging of information between moving vehicles and fixed  roadside infrastructures, co-exist. In order to cope with specific needs of such hybrid scenarios, an adaptation of the  IEEE 802.11 MAC layer, named 802.11p  Wireless Access in Vehicular Environments (WAVE) standard, has been introduced for Vehicular Ad Hoc Networks (VANETs). 
The 802.11p MAC is based on  an  adaptation of the CSMA/CA scheme, called  Enhanced Distributed Channel Access (ECDA) protocol,  to the case of a network whereby traffic with different level of priority, called Access Categories (AC), circulates between wireless nodes. 

\noindent
{\bf Our contribution}. In this paper we present a formal modelling study for the analysis of performances of wireless networks using RTS/CTS based MAC protocols specifically  i) WLANs (i.e. 802.11) and ii)  VANETs (i.e. 802.11p). To this aim we develop   formal models  in terms of Generalized Semi-Markov Process (GSMP) expressed through an high-level stochastic Petri net formalism, namely Stochastic Symmetric  Nets (SSN)~\cite{Chiola:1993:SWC:626501.626829}, which we  analyse through specific performance indicators  defined through  the HASL properties specification language~\cite{BALLARINI201553}.  The models we developed are configurable and allow  for taking into account different scenario including the network dimension,  different incoming traffic conditions, the possible presence of  faults affecting the channel during the transmission of some packet.  For both the 802.11 and 802.11p scenarios we developed and asses KPI expressed in terms of temporal logic specifications. 



\subsection{Related work}
\label{sec:relwork}

Performance analysis of the IEEE 802.11 family of Wi-Fi protocols has been the subject of several research studies~\cite{Bianchi2006,BarbozaASL08,LyaSim2005,KNS02a} each of which considers specific modelling  assumptions concerning e.g.  the traffic model that is considered (i.e. saturated traffic, packets arrival following  a Poisson  law, etc), the presence/absence of errors on the channel  and in case of errors  how  errors are modelled. 
In its pivotal work  Bianchi~\cite{Bianchi2006}, introduced  a simple analytical model that, under given constraints (i.e. finite number of stations and ideal channel condition)  allow for computing the \emph{throughput} of the  IEEE 802.11 Distribution Coordination Function (DCF) for both the Basic Access (BA) and the RTS/CTS versions of the DCF. 
Taking from the two-dimensional discrete-time Markov chain (DTMC) model of Bianchi, many extensions have been considered. In~\cite{AlshanyourA10} a 4D DTMC, inspired by Bianchi's 2D model, has been introduced for considering the case of imperfect channels, that is, taking into account that  transmission over a wireless medium is affected by errors. More specifically in~\cite{AlshanyourA10} the imperfect nature of channels is modelled through a constant \emph{bit error} probability $P_b$ (i.e. the probability that an error occurs during the transmission of  a single bit of data). In~\cite{KNS02a} the BA version of the 802.11 MAC is analysed through  probabilistic model checking  but taking into account only simple modeling assumption (2-nodes network dimension, ideal channel, no specific traffic model). 
\section{RTS/CTS carrier-sensing protocol}
\label{sec:mac}

CSMA/CA MAC schemes are based on the simple idea that a station willing to transmit data packets  has first to gain access to the channel  through a \emph{sensing phase} depending on which the station may either start transmitting, if the  channel has been sensed free along the entire sensing phase, or, if the channel has been sensed busy, refrain from doing so for a randomly chosen duration (backoff). 
Collisions may take place whenever at least two stations ends the sensing phase at the same time, however by employing a randomised backoff delay, the probability of  collisions decreases with the number of successive collisions.  There are two versions of CSMA/CA scheme, the basic access (BA), which we do not consider in this paper, and the Request-to-Send/Clear-to-Send (RTS/CTS).


\begin{figure}[htbp] 
   \centering
   \includegraphics[scale=0.3]{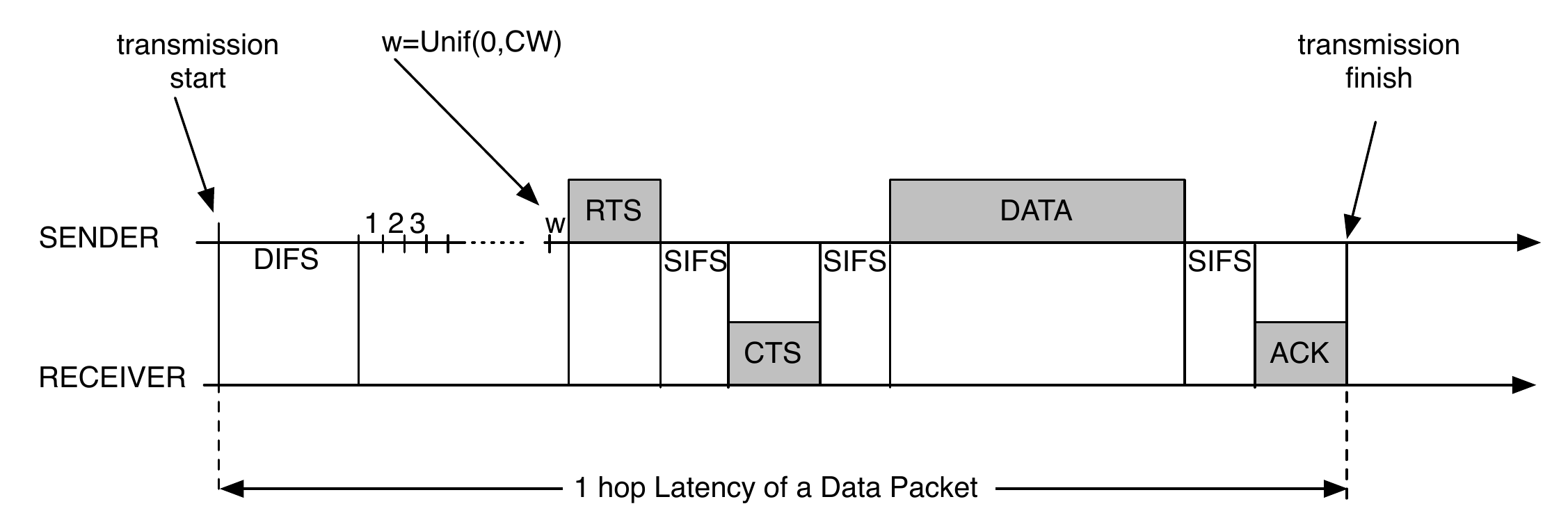} 
   \caption{1-hop carrier-sensing and packets-transmission timing in RTS/CTS 802.11}
   \label{fig:RTSCTStiming}
\end{figure}
\noindent
{\bf 802.11 RTS/CTS:} with the RTS/CTS DCF the transmission of data between a sender and a receiver is regulated by means of a \emph{double handshaking}  scheme which employs  three small-sized, control-packets to regulate the transmission of (larger) data packets (Figure~\ref{fig:RTSCTStiming}). The sender senses the channel for a randomly chosen duration before sending a RTS control packet to the receiver, on reception of which, the receiver, replies with a CTS control packet to the sender (first handshaking). The actual transmission of DATA packets starts as soon as the sender  has received the CTS. Finally when the DATA transmission is over, the receiver acknowledges  the sender with an ACK control packet (second handshaking). The RTS-CTS-DATA-ACK timing sequence is illustrated in Figure~\ref{fig:RTSCTStiming}, which also points out the latency for a successful transmission of a DATA packet over 1-hop. 
To decrease the probability of collisions, the carrier-sensing time $t_{CS}$, also named  {\it Contention Window} (CW), which  is discretely slotted in a finite number of slots of fixed duration $aSlot_t$  ($aSlot_t$ being a parameter of the standard whose value depends on the underlying PHY layer), is randomly chosen. A contention takes place when at least two stations are performing carrier-sensing at the same time. The one that has (randomly) chosen the shortest number of available slots in the CW, wins the contention. The loser, instead, goes into {\it backoff} and increases the dimension of the CW by a power of $2$ (i.e. the new  contention window size $CW'$ is given by  $CW'=(CW_{m}+1)\cdot 2^{bc} -1$ where $bc$, the {\it BackoffCounter}, increases with the number of consecutive unsuccessful transmissions). 
The minimum and maximum size of contention window, respectively $CW_{m}$ and $CW_{M}$, are set by the standard and depends on the PHY layer. For the Frequency Hopping Single Spectrum (FHSS) PHY layer they are equal to $CW_{min}\!=\! 16$ (initial value of $bc=0$) and $CW_{max}\!=\! 1024$ (maximum value of $bc_{max}=6$). Apart from $t_{CS}$, two fixed-length time intervals are also relevant in the RTS/CTS DCF, namely:  {\it Distributed InterFrame Space} (DIFS), the {\it Short InterFrame Space} (SIFS), where $SIFS\!<\!DIFS$ and which are  defined by the PHY layer  in the adopted  networking stack (see Table~\ref{tab:timeparams}).
It should be noted that a collision can take place not only when two contending stations (randomly) pick the same carrier-sensing time (i.e. $t_{CS}$), but also, as pointed out by Heindl {it et al.}~\cite{hege}, because of the existence of the so-called \emph{vulnerable period}, which accounts for three factors: the time of radio waves propagation through the medium ($aAirPropagationTime$), the time a station takes for accessing the medium ($aCCATime$) and the time a station takes for switching from receiving to transmitting mode($aRxTxTurnaroundTime$). As a consequence the duration of  $aSlot_t$ is  set by the IEEE standard to a value larger than the \emph{vulnerable period} (i.e. the value $aSlot_t$   depends on the PHY-layer dependent parameters $aCCATime$ and $aRxTxTurnaroundTime$ plus the negligible $aAirPropagationTime$). 

\noindent
{\bf 802.11p RTS/CTS with priority classes:}
The 802.11p MAC is based on an extension of  the CSMA/CA  RTS/CTS scheme, called Enhanced Distributed Channel Access (ECDA), which supports  4 priority levels, called Access Categories (AC), for   data traffic,  namely: Background (AC\_BK), i.e. the  lowest priority, Best effort (AC\_BE), Video (AC\_VI) and Voice (AC\_VO), i.e. the highest priority. EDCA  is designed so that higher-priority traffic is more likely (than lower-priority) to be be granted access to the shared medium and successfully being transmitted. In practice this is obtained by associating  each AC with a \emph{contention window} (CW) whose size is inversely proportional to the corresponding priority level. Therefore, for example, the CW for AC\_BK (lowest priority) has to be larger than that of AC\_VO (highest priority). Table~\ref{tab:contentionwindowpar} depicts the CW characterisation for the  ACs of ECDA, where  \texttt{aCWmin} and \texttt{aCWmax} are parameters set by the 802.11p protocol. One can notice that  AC\_VO is the dominant AC as its CW is not only the minimal one but also  it does not overlap to the CW of any other AC. 


\section{Background: stochastic symmetric Petri nets}
\label{sec:background}
To carry out the performance modelling study of 802.11 protocols   we used a \emph{coloured} stochastic Petri net formalism, namely the Stochastic Symmetric  Net (SSN)~\cite{Chiola:1993:SWC:626501.626829}, to model networks using 802.11/802.11p MAC, and we applied  the Hybrid Automata Specification Language (HASL)~\cite{BALLARINI201553} formalism (through the \cosmos\ statistical model checker~\cite{BDDHP-qest11}) to assess  relevant performance indicators against the SSN models. For the sake of space we only give a very succinct overview of the SSN and HASL formalisms referring the reader to the literature for a detailed  treatment.

\noindent
{\bf Stochastic Symmetric Nets}. 
Like any  Petri net formalism an SSN model is a bi-partite graph consisting of \emph{place} nodes (circles) and \emph{transition} nodes (bars). Places  may contain (countably many) \emph{tokens} and are connected to transitions through arcs which are labelled with \emph{arc functions}. An SSN model describes a continuous-time, discrete-state stochastic process whose states correspond with the possible \emph{markings} of the SSN (a marking gives the content of each place of the SSN). The peculiarity of SSN models is that tokens may be associated with information, hence they may have different \emph{colours}, instead of being all  indistinguishable as with non-coloured PN formalisms. Therefore places and transitions of an SSN are associated with a \emph{color domain} (CD) built from
elementary types called \emph{color classes} ($\{C1, . . . , Cn\}$) with $cd(p)$, resp. $cd(t)$, denoting the CD of place $p$, resp. transition $t$ (see Table~\ref{tab:colordomain} and Table~\ref{tab:colorclasses} for the CDs of 802.11p SSN model). SSN color classes are finite, non empty and disjoint sets, they may be ordered (in this case a successor function is defined on the class, inducing a circular order among the elements in the class), and may be partitioned into (static) subclasses. An SSN transition may be immediate (drawn as a thin filled in bar) or timed which means its firing  delay is sampled (the moment it gets enabled) from a probability distribution that may be \emph{exponential} (thick empty bar) or \emph{general} (thick filled in bar). A coloured transition may be associated with a \emph{guard}, i.e. a boolean-valued expression built on top of \emph{colored variables} by means of the following  basic predicates: $x\! =\! y$, $x\!\in\! \mathit{subclass}$, $d(x)\! =\! d(y)$ where $x$ and $y$ are variables of the transition with same type, and $d(x)$ denotes the static subclass $x$ belongs to.A valid \emph{transition binding} is an assignment of values to its variables, satisfying the predicate expressed by the \emph{guard}. A pair (transition,binding) is called \emph{transition instance}. Each arc connecting a place $p$ and a transition $t$ is labeled with an expression denoting a function $\mathit{arcf : cd(t) \to Bag(cd(p))}$ where $Bag(A)$ is the set of all possible multisets that may be built on set $A$. The valuation of $\mathit{arcf}$ given a legal binding of $t$ gives the multiset of colored tokens to be withdrawn from (in case of input arc) or to be added to (in case of output arc) the place connected to that arc upon firing of such transition instance. The arc expressions in SSNs are built upon a limited set of primitive functions whose domains must be color classes. Typically an arc expression is a linear combination of function tuples (denoted $\langle f1, \ldots ,fn\rangle$), and each element of a tuple is either a projection function, denoted by a variable in the transition color domain (e.g. $sa$ and $sb$ in the tuple $\langle sa,sb,p,pt\rangle$   appearing as the labelling in several arcs of the SSN of Figure~\ref{fig:80211pRTSCTS}), a successor function, denoted $\mathit{x\!+\!+}$ were $x$ is a variable  whose type is an ordered class; a constant function, denoted $\mathit{C_i.All}$ returning all elements of (sub)class  $C_i$; a complement function denoted $\mathit{C_i.All}\!-\! x $ where $x$ is a variable of type $C_i$.  
The dynamics  of an SSN model is defined in terms of transition instances \emph{enabling} and \emph{firing}: a transition instance is \emph{enabled} if the marking of all of its input places is \emph{compatible} with it (i.e. if the marking enables the transition instance). Upon \emph{firing} a transition instance modifies the state of the SSN by removing (resp. adding) tokens from its input (resp. output) places. For example, w.r.t. the SSN in Figure~\ref{fig:80211pRTSCTS}, assuming colour classes $\mathit{St\!=\!\{st1,st2\}}$ and $\mathit{Pr\!=\!\{pr1\}}$, the initial marking of place $\mathit{Idle}$ (colour domain  $\mathit{St\!\times\! Pr}$) being $\mathit{\langle St, Pr\rangle \!=\!\{\langle st1,pr1\rangle,\langle st2,pr1\rangle\}}$  enables two instances of  timed transition $\mathit{PacketArrival}$ (given its guard $sa\!\neq \! sb$), namely  $\mathit{\langle st1,st2,pr1,rts\rangle}$ $\mathit{\langle st2,st1,pr1,rts\rangle}$. The firing of the first instance consumes $\mathit{\langle st1,pr1\rangle}$ from $\mathit{Idle}$ and adds $\mathit{\langle st1,st2,pr1,rts\rangle}$ to $\mathit{Sense}$.

\vskip -10ex
\section{Modelling  of 802.11/802.11\lowercase{p} networks}
\label{sec:prismperfanalysis}
To analyse the performance of two versions of the protocol we developed two SSN models, one modelling a network  whose stations use the 802.11 MAC, the other where stations that use the 802.11p MAC. The models we developed are based on the following  assumptions: {\bf 1) Clicque network topology}: the network consists of $N$  stations arranged in  a clique (i.e.  every station can overhear every other station).  {\bf 2) Traffic direction}: each station  has a unique target station to which it addresses    its incoming traffic. {\bf 3) Incoming traffic}:  each station is either under  a) a \emph{saturated  regime} (i.e. a packet ready to be transmitted is invariably present) or  b) its incoming traffic is given by an   Poisson process with parameter $\lambda$.  {\bf 4) Perfect/Imperfect channel}: the wireless medium  is supposed to behave either a) as a \emph{perfect channel} (i.e. transmitted  data are never  affected by errors) or  b) as an \emph{imperfect channel} with some error probability (details below).  {\bf 5) Vulnerable period}:  the radio device of each station is supposed to exhibit a certain delay for switching between transmission/reception mode. The duration of such period is called the \emph{vulnerable period} and is one source  of traffic collisions within the network.  \\
\noindent
\emph{Channel error model.} 
To model the possibility that transmissions undergo errors due to the channel we adopted the burst-noise binary channel model~\cite{Gilbert60}. 
A burst noise channels may be subsumed  by a two-states Markov chain where one state ($G$ as in good) represents the absence of noise while the other ($B$  as in burst errors) represents the presence of an error spike which affects the channel impeding a transmission to correctly take place. We equipped our Petri nets models with an implementation of such burst-noise binary channel model.

\subsection{SSN models  of 802.11/802.11\lowercase{p} networks}
\label{sec:perfanalysis80211p}



We developed two SSN models one for 802.11 networks the other for 802.11p's. For the sake of space we only present the 802.11p version of the model, which is somehow a generalisation of the 802.11's.   The 802.11p SSN model relies on the definition of a number of color classes (Table~\ref{tab:colorclasses}) and color domains (Table~\ref{tab:colordomain}) that are used to characterise the type of the various elements (places and tranisitions).  The  model consists essentially of  three parts: the \emph{802.11p core module}, representing the prioritised carrier-sensing and RTS/CTS handshaking mechanism, the \emph{backoff module}, representing the behaviour of a station in case of a collision and the \emph{medium module} representing the state of the wireless channel. We present these 3 modules separately bearing in mind that they are part of the same SSN model. 

\begin{figure*}
\begin{center}
\includegraphics[scale=.3]{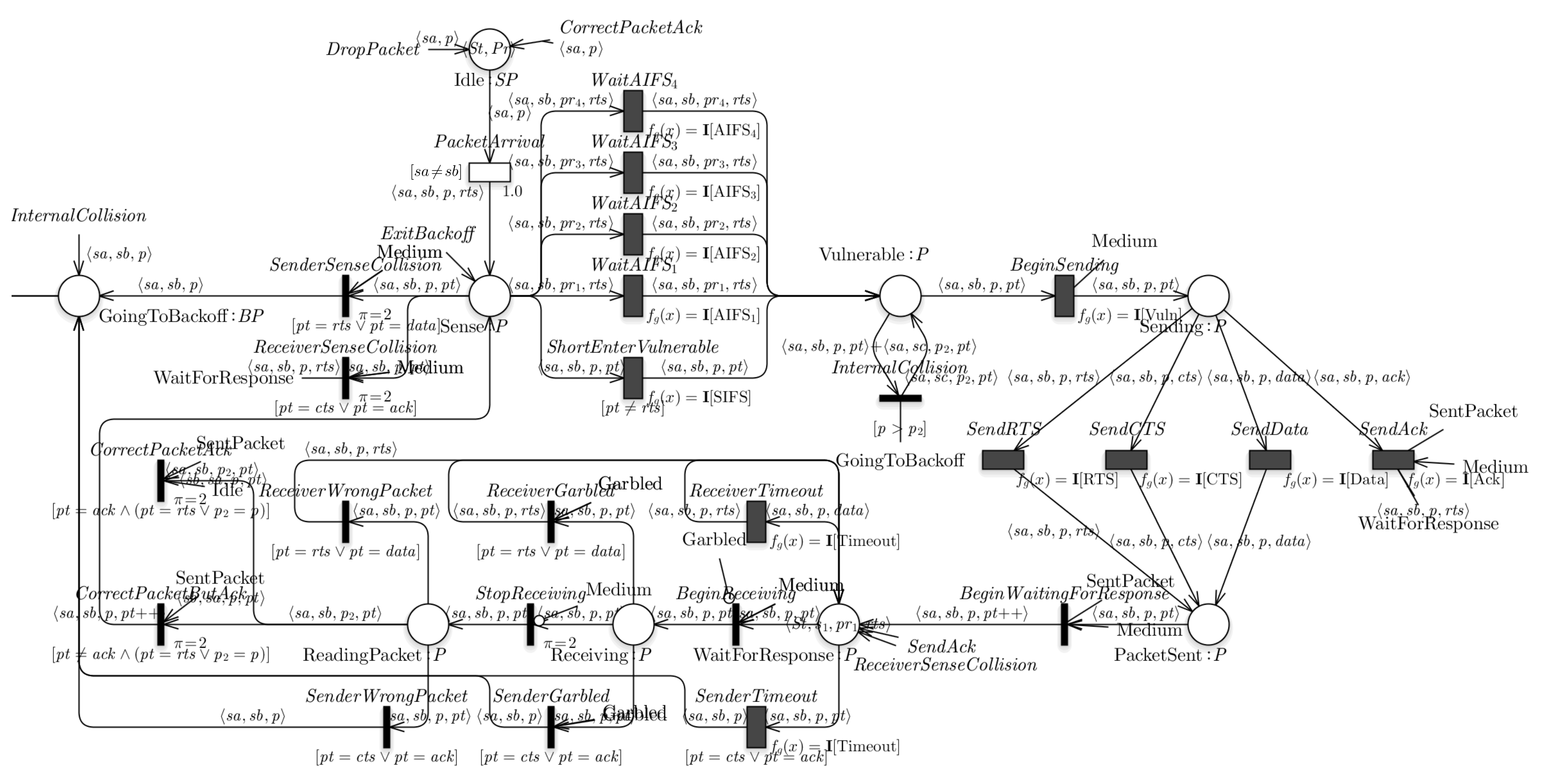}
\caption{SSN model of the 802.11p prioritised RTS/CTS scheme}
\label{fig:80211pRTSCTS}
\end{center}
\end{figure*}

\noindent
\paragraph{802.11p core SSN module}
Figure~\ref{fig:80211pRTSCTS} depicts the portion of the SSN model representing the traffic generation and  prioritised  RTS/CTS handshaking protocol. 
It consists of three main parts: the generation of  incoming data traffic (timed transition $\mathit{PacketArrival}$), the carrier-sensing phase (timed transitions $\mathit{WaitAIFS}_i$, $\mathit{WaitSIFS}$), the packets transmission  phase (timed transitions $\mathit{sendRTS}$, $\mathit{sendCTS}$, $\mathit{sendDATA}$, $\mathit{sendACK}$), the handling of packets received/overheard by a station (the sub-net in between places $\mathit{WaitForResponse}$, $\mathit{Receiving}$ and $\mathit{ReadingPacket}$). Let us describe the various parts of the model. Place $\mathit{Idle}$ (color domain $\mathit{SP}$) is initially filled up with tokens corresponding with all pairs from $\mathit{St}\!\times\!\mathit{Pr}$, representing that each station may generate incoming traffic with any level of priority. The firing of the outgoing timed (exponentially distributed) transition $\mathit{PacketArrival}$ consumes a $\langle sa,p\rangle$ token and produces the $\langle sa,sb,p,rts\rangle$   token (added to place $\mathit{Sense}$) indicating that station $sa$ is ready to send an $rts$ request with priority $p$ to station $sb$ ($sa\!\neq\! sb$). A token $\langle sa,sb,p,pt\rangle$   in place  $\mathit{Sense}$ (representing an ongoing transmission of a packet of type $pt$ and priority $p$ from station $sa$ to $sb$) is then moved either to i) place $\mathit{Vulnerable}$ through either one of the four mutually-exclusive timed (deterministic) transition  $\mathit{WaitAIFS}_i$ (only that corresponding to the actual priority value $p$ is enabled), if  packet type $pt\!=\!rts$ and the medium remains free for  $\mathit{AIFS}_i$ time units, or through $\mathit{WaitSIFS}$ if packet type $pt\!\neq\! rts$ and medium stay free for $\mathit{SIFS}$, or if the medium gets occupied in the meantime ii) to $\mathit{GoingToBackoff}$  (through immediate transition $\mathit{SenderSenseCollision}$) if $sa$ is the sender of an $rts$ or $data$ packet or  iii) to $\mathit{WaitForResponse}$ (through immediate transition $\mathit{ReceiverSenseCollision}$) if $sa$ is a station responding with either a $cts$ or $ack$ to $sb$. A token $\langle sa,sb,p,pt\rangle$ from place $\mathit{Vulnerable}$ moves to $\mathit{Sending}$ (after $\mathit{Vuln}$ time units through timed  transition $\mathit{BeingSending}$), and in so doing it adds one (uncoloured) token to $\mathit{Medium}$ indicating that the number of transmitting stations has   increased. Observe that the immediate transition $\mathit{InternalCollision}$ (which is also consuming tokens from $\mathit{Vulnearble}$) deals with the case of   competing packets sent,  with different level of priorities,  by a common station $sa$: 
in this case  only the packet with the highest priority stays in $\mathit{Vulnerable}$, while packets with lower priority are  moved to   $\mathit{GoingToBackoff}$ (i.e. highest priority packet wins the internal competition by ``overtaking'' lower priority packets). From place $\mathit{Sending}$ a token  $\langle sa,sb,p,pt\rangle$ moves to   $\mathit{PacketSent}$ if the packet type is either $rts$, $cts$ or $data$, after a  delay corresponding with the kind of packets, and then, with no delay, it is further moved to $\mathit{WaitForResponse}$ through immediate transition $\mathit{BeginWaitingForResponse}$ which while removing a token from $\mathit{Medium}$ (hence decreasing the occupation of the wireless channel) and storing the $\langle sa,sb,p,pt\rangle$  in place $\mathit{SentPacket}$ for later retrieval, also updates the packet type of token  $\langle sa,sb,p,pt\rangle$ to $\langle sa,sb,p,pt\!+\!+\rangle$, indicating that station $sa$ is now ready to wait for a reply message (of type $pt\!+\!+$, i.e. $cts$ in reply to $rts$ or $ack$ in replay to $data$) from $sb$. 
Place $\mathit{WaitForResponse}$ is initialised with marking $\langle St,s1,pr1,rts\rangle$ indicating that each station of $St$ is, initially, ready to get engaged into replying to an $rts$ request ($s1$ and $pr1$ being irrelevant at this stage). As the medium gets occupied but not garbled all tokens in $\mathit{WaitForResponse}$ are moved (with no delay) to $\mathit{Receiving}$ and from there, as soon as the channel gets free (and assuming it hasn't been garbled before) they  move on to $\mathit{ReadingPacket}$. Thus, at the end of a transmission phase, place  $\mathit{ReadingPacket}$ contains all tokens representing all possible combinations of responses that all stations may be engaged in. On completion of the transmission corresponding to token  $\langle sb,sa,p,pt\rangle$ being added to $\mathit{SentPacket}$ the corresponding token $\langle sa,sb,p_2,pt\rangle$, if present,  is consumed from $\mathit{ReadingPacket}$ (through either one of the prioritised transitions $\mathit{CorrectPacketButAck}$ or $\mathit{CorrectPacketAck}$) representing the creation of the response, by the destination station $sa$, to  the transmitted packet $pt$. Notice that $\mathit{CorrectPacketButAck}$  triggers the response to an $\mathit{rts}$, $\mathit{cts}$ or $\mathit{data}$  packet   by adding the token $\langle sa,sb,p,pt\!+\!+\rangle$ to $\mathit{Sense}$ hence moving to the next step of the RTS/CTS protocol for the processed packet. Conversely $\mathit{CorrectPacketAck}$ represents the end of the RTS/CTS handshaking hence it puts back a $\langle sa,p\rangle$ token to $\mathit{Idle}$ which restart  the cycle for the transmission of a priority level $p$ packet by station $sa$.  On the other hand all  tokens $\langle sa,sb,p,pt\rangle$ that remain in $\mathit{ReadingPacket}$ after a response to the transmitted packet has been treated (through either $\mathit{CorrectPacketButAck}$ or $\mathit{CorrectPacketAck}$)   are either put back to $\mathit{WaitForResponse}$ (if they represent a receiver, i.e. transition $\mathit{ReceiverWrongPacket}$) or  they are moved to $\mathit{GoingToBackoff}$ (if they represent a sender that  was expecting either a   $\mathit{cts}$ or an $\mathit{ack}$ response to a previosuly sent packet $\mathit{rts}$ or $\mathit{data}$). 

\begin{figure*}
\begin{center}
\includegraphics[scale=.3]{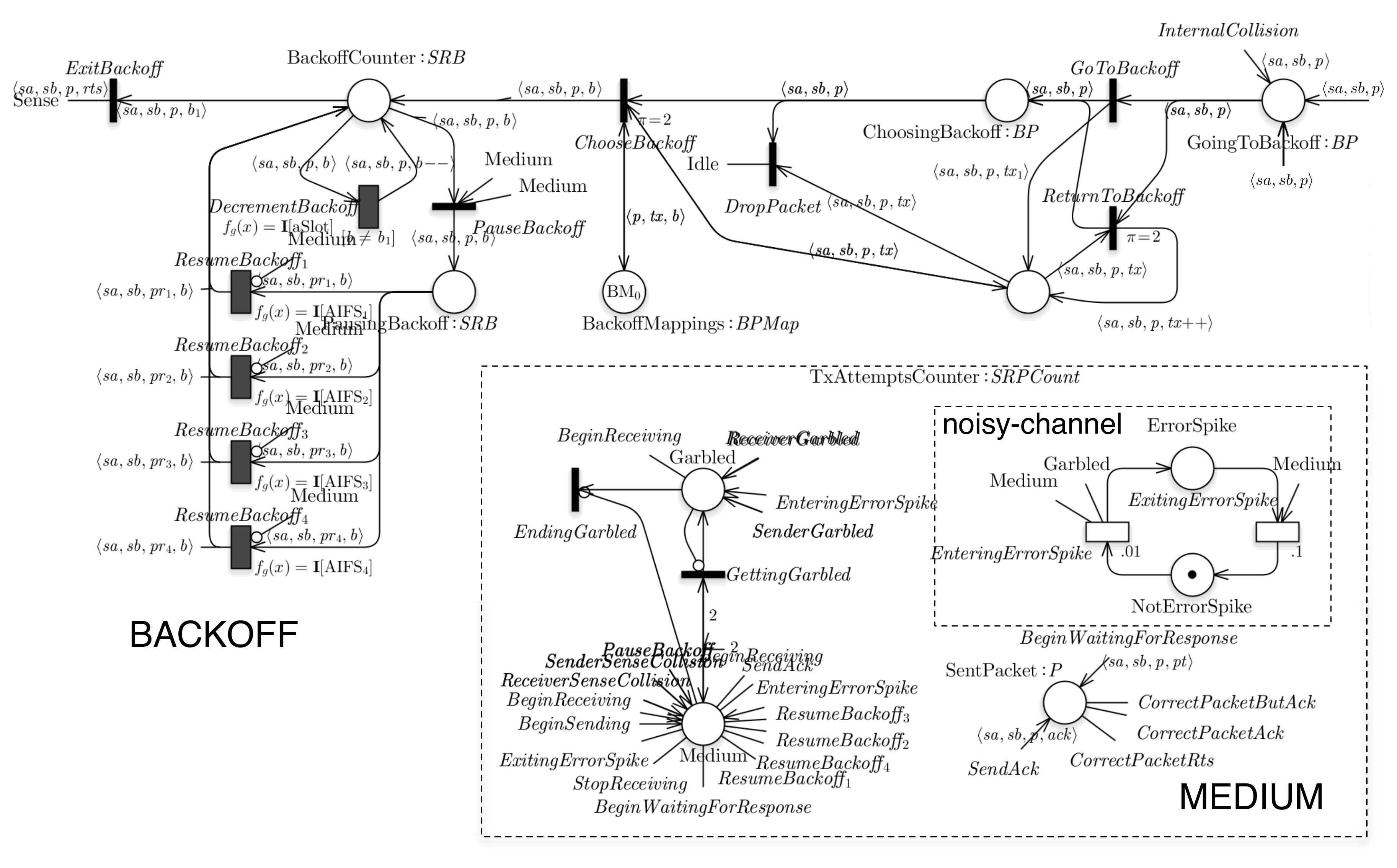}
\caption{SSN model of the 802.11p backoff mechanism (top left) and the medium (bottom right).}
\label{fig:80211pbackoff}
\end{center}
\end{figure*}

\noindent
\paragraph{802.11p backoff SSN module}
With the 802.11p DCF the selection of the \emph{backoff} duration, i.e. the selection of the number of  \emph{time slots} a packet has to wait before starting a new transmission attempt, depends on the priority class of the packet as well as on the backoff round, i.e. the number of times a packet has unsuccessfully went through the carrier-sensing phase without winning the contention. Therefore a model of 802.11p backoff procedure must be equipped with necessary  means to take  into account the priority level of each packet hence  the size of the contention window for each  priority level.
Figure~\ref{fig:80211pbackoff} (top left) depicts the SSN subnet  representing the randomised selection of the backoff delay for a packet of a given priority level $p$. A backoff begins when a token $\langle sa,sb,p\rangle$ is added to place $\mathit{GoingToBackoff}$ (color domain $\mathit{BP}$) indicating that something went wrong (e.g. a collision or a lack of handshaking) during the transmission of  a priority $\mathit{p}$ packet from station $\mathit{sa}$ towards $\mathit{sb}$. The first step then is the updating of the backoff round counter  which boils down to adding of a token $\langle sa,sb,p,tx \rangle$  to place $\mathit{TxAttemptsCounter}$ (whose  color domain is $\mathit{SRPCoun}t$ and where  variable $\mathit{tx}\!\in\! \mathit{TxCount}$ stores the backoff counter)  through either   transition $\mathit{GoToBackoff}$ (enabled if the failed transmission attempt is the first one, hence initialising $\mathit{tx}$ to color $\mathit{tx_1}$)  or through transition $\mathit{ReturnToBackoff}$ (enabled by any further failed transmission attempt and increasing $\mathit{tx}$ by 1). 
Once a token  $\langle sa,sb,p \rangle$ enters place $\mathit{ChoosingBackoff}$ having, in the process, also added (or updated) a  $\langle sa,sb,p,tx \rangle$  token to  place $\mathit{TxAttemptsCounter}$, either one between the  mutually exclusive transitions $\mathit{ChooseBackoff}$ or $\mathit{DropPacket}$ transition is enabled. Transition $\mathit{ChooseBackoff}$ realises the probabilistic selection of the backoff counter value $b$, corresponding to the priority level $p$ of the \emph{backoffing} packet, by adding token $\langle sa,sb,p,b\rangle$ to place $\mathit{BackoffCounter}$. Such a  selection is achieved through  randomly choosing (through  the test arc between  transition $\mathit{ChooseBackoff}$ and place $\mathit{BackoffMappings}$) a token   $\langle p,tx,b \rangle$ among those specified by the invariant marking  of  $\mathit{BackoffMappings}$ (Table~\ref{tab:marking}). 
The marking of $\mathit{BackoffMappings}$ associates with each priority level $\mathit{pr_i}$ and re-transmission attempt $\mathit{tx_j}$ the corresponding CW expressed as  the union of color subclasses $\cup_{k=1}^{k'} bs_k$ (where $k'$ is a function of $pr_i$ and $tx_j$). 
Therefore, for example,  for the first re-transmission attempt (i.e. $\mathit{tx=tx_1}$) of a priority  $p_1$ packet  the backoff value $b$ is chosen randomly as $b\in\ \mathit{bs_1}=\mathit{\{b1,\ldots ,b4\}}$ (which corresponds with the narrowest CW i.e. the AC\_V0  category) as the initial marking $\mathit{BM_0}$ contains  tokens $<pr1,tx1,bs1>$.  Similarly for the second re-transmission attempt (of a priority  $p_1$ packet) $b$ is chosen randomly as $b\in\ \mathit{bs_1}\cup \mathit{bs_2}=\mathit{\{b1,\ldots ,b8\}}$ as marking $\mathit{BM_0}$ contains  tokens $<pr1,tx2,bs1+bs2>$ and so on.  Once a $\langle sa,sb,p,b\rangle$ token is added to place $\mathit{BackoffCounter}$ (i.e. the backoff delay $b$ is selected) the actual backoff (sensing) period  begins: $b$ is decremented every $\mathit{aSlot}$ time units (deterministic transition $\mathit{DecrementBackoff}$) however, as soon as the medium gets occupied, the decrement of $b$ is suspended by moving token $\langle sa,sb,p,b\rangle$ from  $\mathit{BackoffCounter}$ to $\mathit{PausingBackoff}$ (immediate transition $\mathit{PauseBackoff}$).  Token $\langle sa,sb,p,b\rangle$ is moved back to  $\mathit{BackoffCounter}$ only after the channel got freed and stayed free for $\mathit{AIFS_i}$ time units (where $i\!\in\!\{1,2,3,4\}$ denotes the priority level $p$  of the  backoff-ing packet). The backoff  for a token $\langle sa,sb,p,b\rangle$ ends as soon as the backoff counter has reached 1 (i.e. $b\!=\!b1$): at this point token $\langle sa,sb,p,b1\rangle$ is removed from $\mathit{BackoffCounter}$ (immediate transition $\mathit{ExitBackoff}$) and $\langle sa,sb,p,rts\rangle$ is pushed back to $\mathit{Sense}$ which represent the re-start of the transmission procedure for station $\mathit{sa}$ to sent a priority $p$ packet to $\mathit{sb}$.
\paragraph{802.11p medium SSN module}
Figure~\ref{fig:80211pbackoff} (bottom right) shows the part of the SSN model  representing the shared channel. It consists of 2 uncoloured places: $\mathit{Medium}$,   whose marking corresponds with the number of transmitting stations, and  $\mathit{Garbled}$,  which contains a token whenever a collision has taken place or an error spike has occurred on the channel.  Transition $\mathit{GettingGarbled}$ sets the state of the medium to \emph{garbled} (adding a token in place $\mathit{Garbled}$) as soon as  $\mathit{Medium}$ contains at least 2 tokens (i.e. at least two stations are transmitting).   Transition $\mathit{EndingGarbled}$ ends the \emph{garbled} state of the medium by removing a token from $\mathit{Garbled}$ as soon as the $\mathit{Medium}$ is emptied.  Place $\mathit{SentPacket}$ (domain $P$) stores the information relative to an ongoing transmission in terms of   a token $\mathit{\langle sa,sb,p,pt\rangle}$ where  $sa$ is the sender station, $sb$ the receiver ,$p$  the priority level $p$ and  $tp$ the packet type). The medium SSN is also equipped with a subnet for modelling the presence of error spikes affecting an ongoing transmission. To study the effect of a noisy channel on the network performances it suffices that place $\mathit{NotErrorSpike}$  initially contains a token in which case transition $\mathit{EnterinErrorSpike}$ reproduce the occurence of errors on the channel (following an Exponential distribution with configurable rate, $0.01$ in the picture), by adding a token in $\mathit{Medium}$ (hence triggering a collision if a transmission is going on). The end of an error-spike is modelled by transition $\mathit{ExitingErrorSpike}$ (also Exponentially distributed with a configurable rate).

\section{analysis of 802.11/802\lowercase{p} models}
\label{sec:cosmosmodel}



\begin{figure}[htbp]
\begin{center}
\begin{tikzpicture}[scale=1.5]
	\node at (1.200000,9.200000) [draw,rounded corners] (n0) {l1;$\dot{t}=1$};
	\node at (-0.300000,9.200000) [] (n1) {};
	\node at (3.000000,9.200000) [rounded corners,draw,accepting] (n2) {l2};
	\path[draw,draw,-latex] (n1)  -- node[above] {\#;;t=0} (n0);
	\path[draw,-latex] (n0)  -- node[above] {\#;t=T} (n2);
	\path[draw,-latex] (n0)  .. controls (0.600000,8.400000) and (1.800000,8.400000) ..   node[below] {CorrectPacketAck(p);t<T; Throughtput[p]++;} (n0);
	\path[draw,-latex,loop] (n0)  .. controls (0.600000,10.000000) and (1.800000,10.000000) ..   node[above] {All - CorrectPacketAck;t<T;} (n0);
\end{tikzpicture}
\end{center}
\caption{The LHA for measuring the throughput}
\label{fig:lha1}
\end{figure}
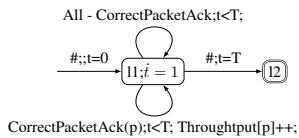

We analysed the  performances of both the 802.11 and 802.11p  network models~\footnote{SSN models and HASL properties used to run  such experiments on \cosmos\  are available at \url{https://sites.google.com/site/pballarini/models.}} by means of the \cosmos\ statistical model checker~\cite{BDDHP-qest11, cosmos}.  We  considered a number of   key performance indicators (KPI) including  1) the {\bf throughput} (THR) of a network station (i.e. the number of successfully transmitted  packets per time unit); 2) the {\bf busyTimeRatio} (BTR, i.e. the ratio between the time the channel is occupied by some transmitting station   and the total   operation time of the network).  We encoded such KPIs in terms of HASL specifications for SSN models~\cite{Amparore:2013:SVH:2486092.2486124}  and used them to run a number of experiments to assess   the impact that 1) a faulty channel, 2) the incoming traffic,   3) the network dimension have on the KPIs. In the remainder we present  an excerpt of the results obtained. Figure~\ref{fig:lha1} shows an example of an HASL specification (a linear hybrid automaton, LHA) we used to assess the THR for   priority $p$ traffic on the 802.11p model. 
The LHA in Figure~\ref{fig:lha1} has an initial ($l_1$) and a final ($l_2$) location and uses a clock ($t$) plus a counter of successfully terminated transmissions for a priority $p$ packet ($Throughput[p]$). On processing a trajectory $Throughput[p]$ is incremented each time  an instance of the $\mathit{CorrectButAck}$ transition (Figure~\ref{fig:80211pRTSCTS}) with the corresponding priority $p$ level occur (which coincides with the termination of  the  transmission of a  priority $p$ packet). The LHA ends measure at time $T$ and the value of THR is obtained as $Throughput[p]/T$.

\begin{figure*}
\begin{center}
\subfigure[Throughput 802.11]{\includegraphics[scale=.2]{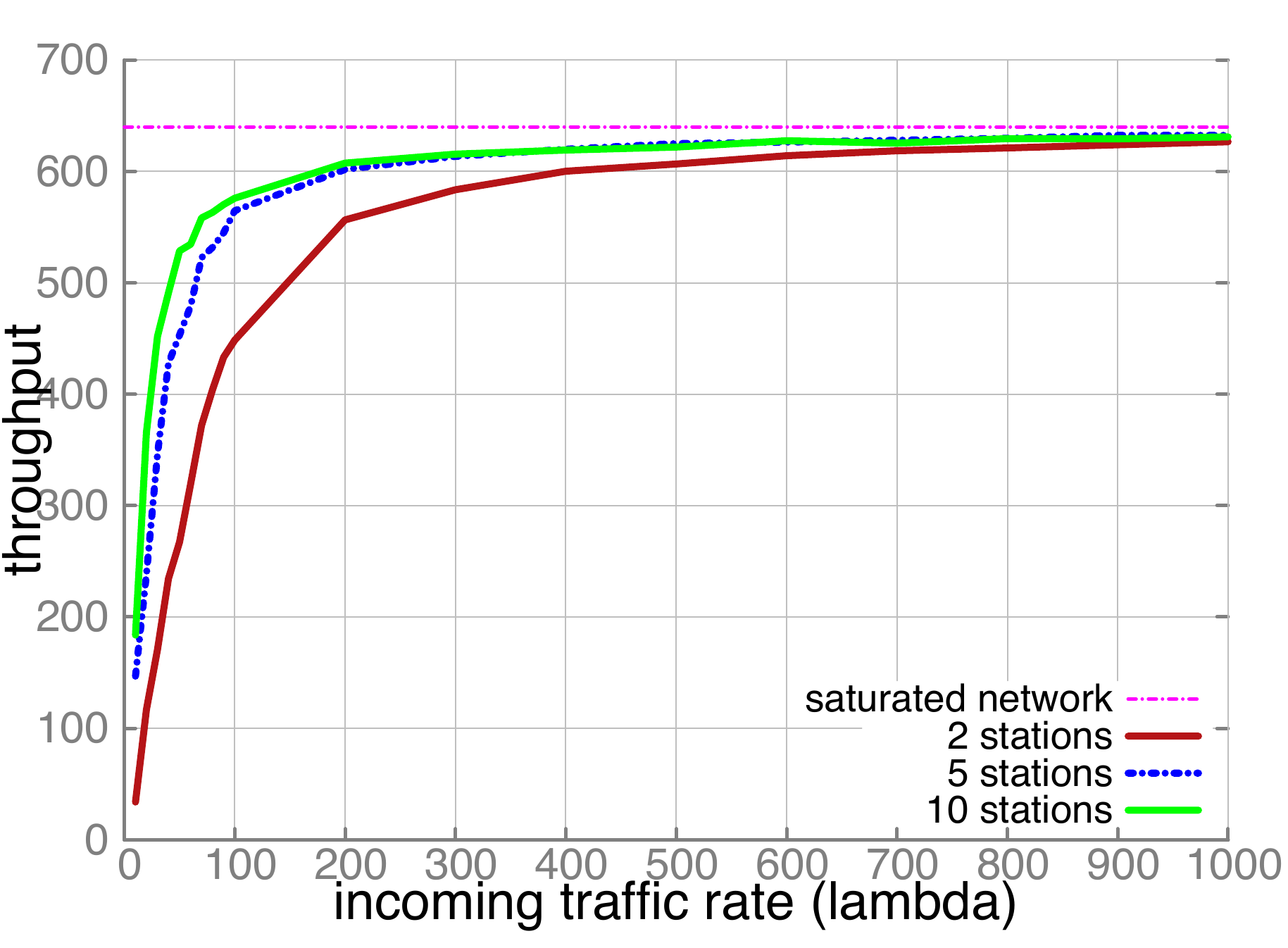}\label{fig:THR_80211}}
\subfigure[BusyTimeRatio 802.11]{\includegraphics[scale=.2]{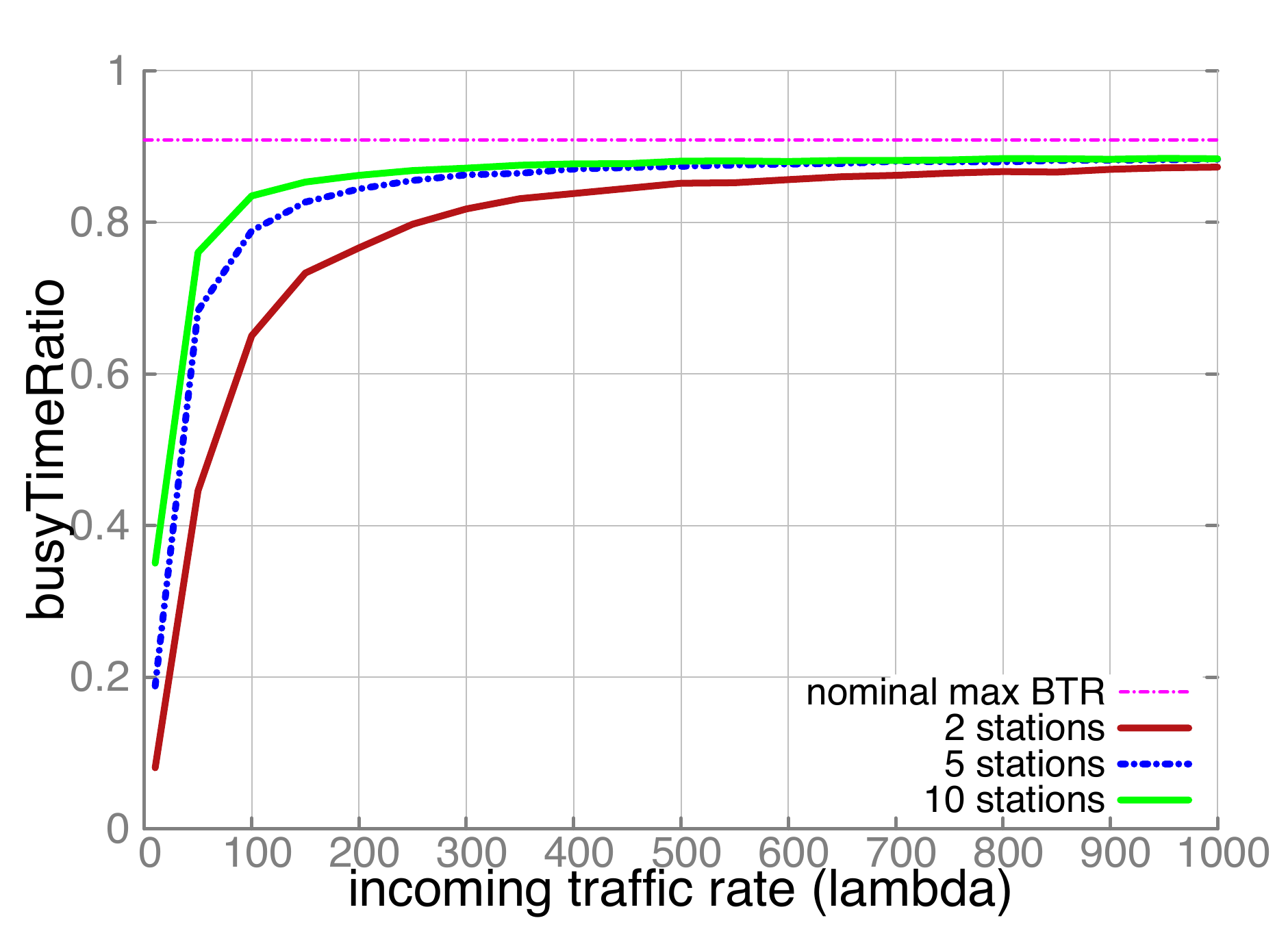}\label{fig:BTR_80211}}
\subfigure[saturated traffic 802.11]{\includegraphics[scale=.2]{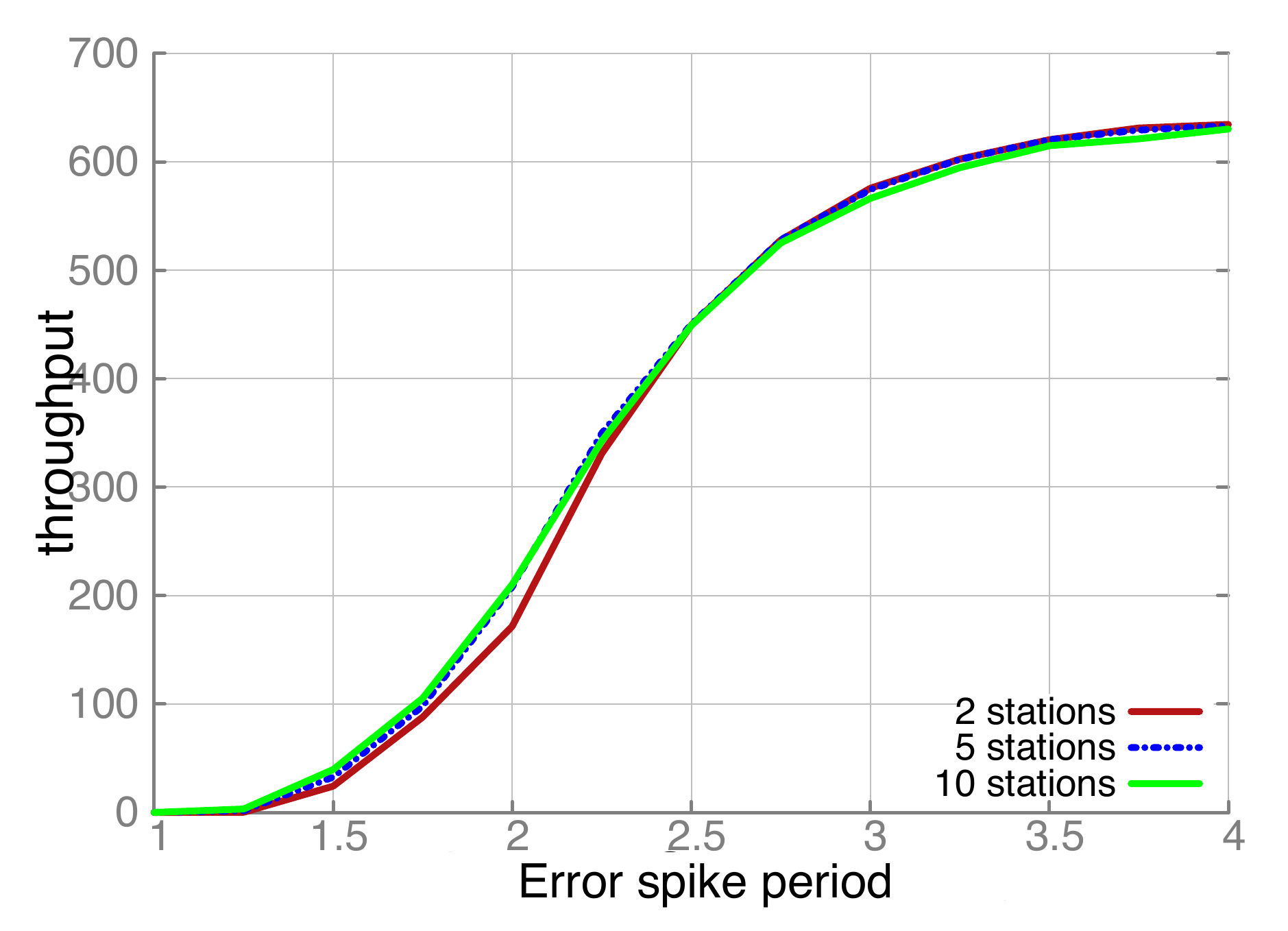}\label{fig:THR_ERR_80211}}
\subfigure[\small 2 stations, non-saturated traffic 802.11]{\includegraphics[scale=.2]{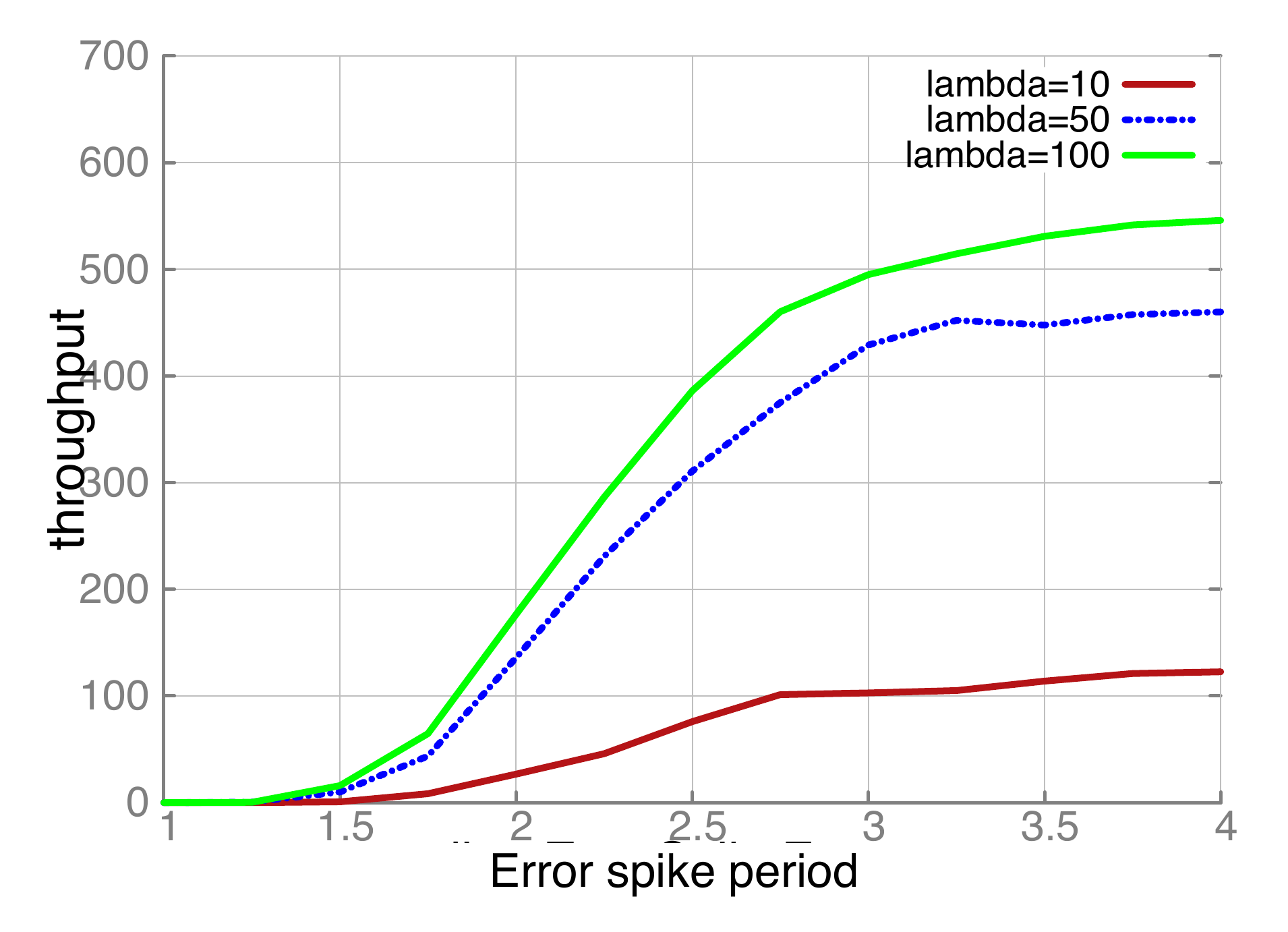}\label{fig:THR_ERR2_80211}}
\subfigure[Throughput 802.11p]{\includegraphics[scale=.15]{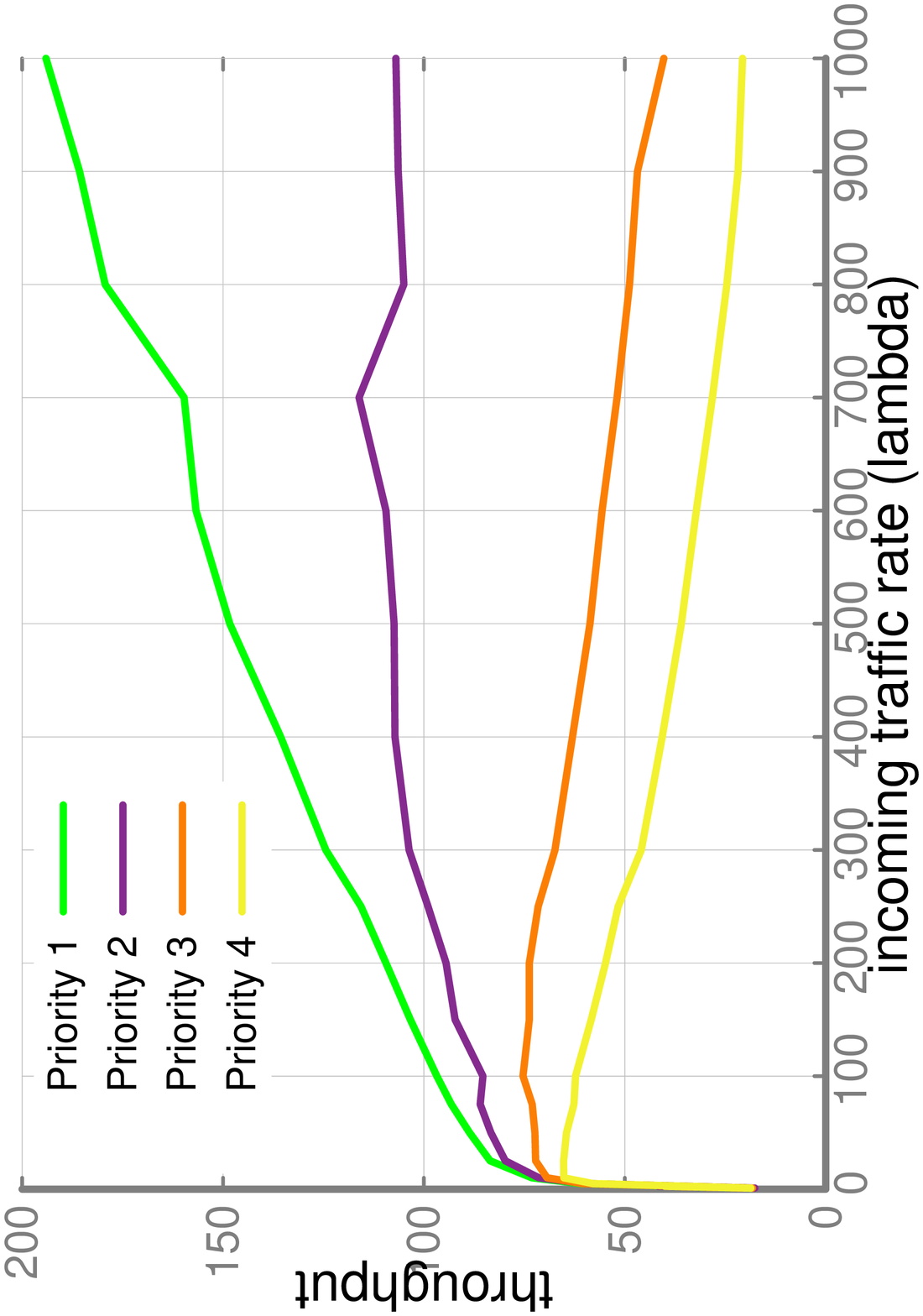}\label{fig:THR_80211p}}
\hskip -4ex
\subfigure[BusyTimeRatio 802.11p]{\includegraphics[scale=.15]{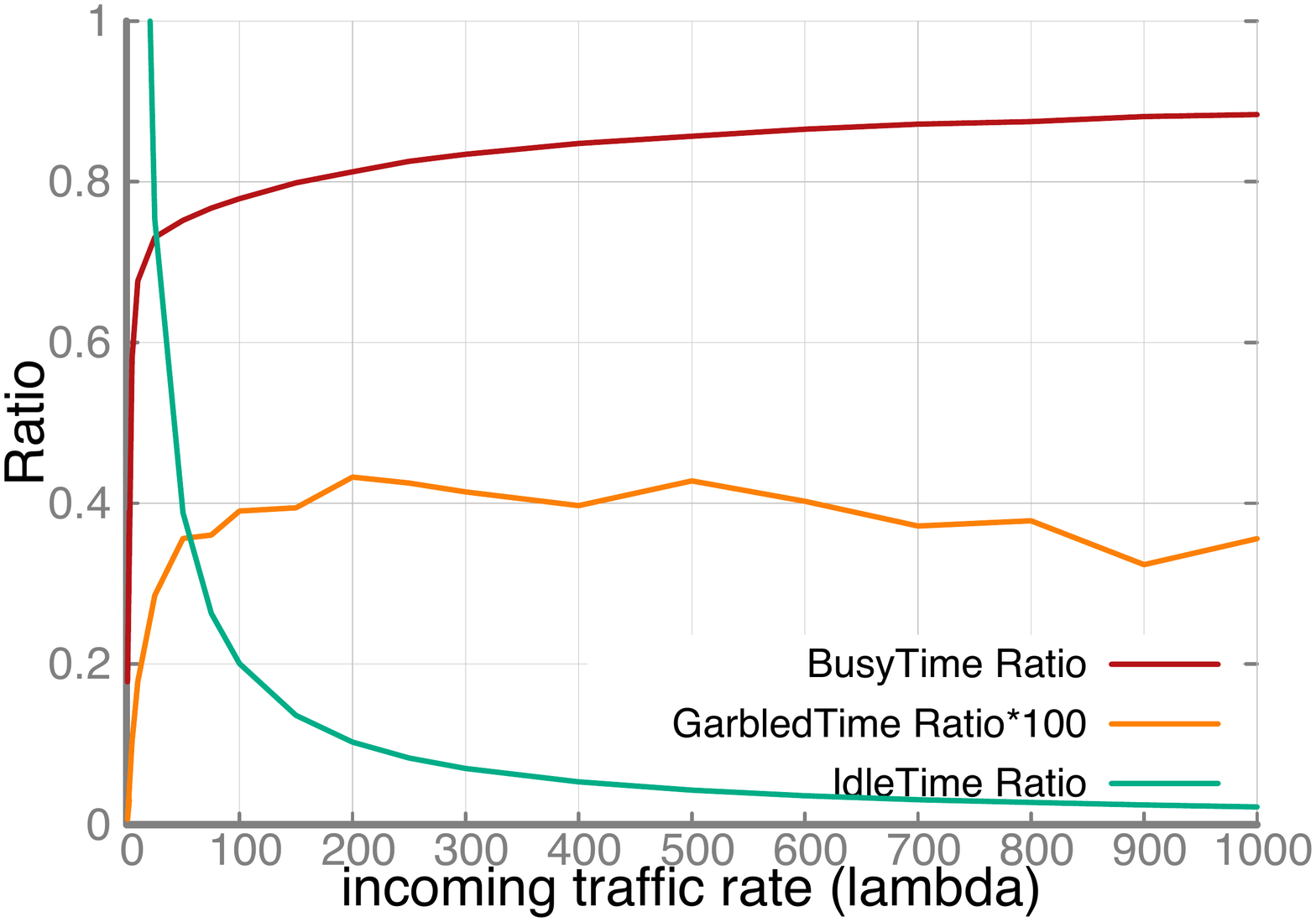}\label{fig:BTR_80211p}}
\hskip -4ex
\subfigure[traffic rate $\lambda\!=\! 100$ 802.11p]{\includegraphics[scale=.15]{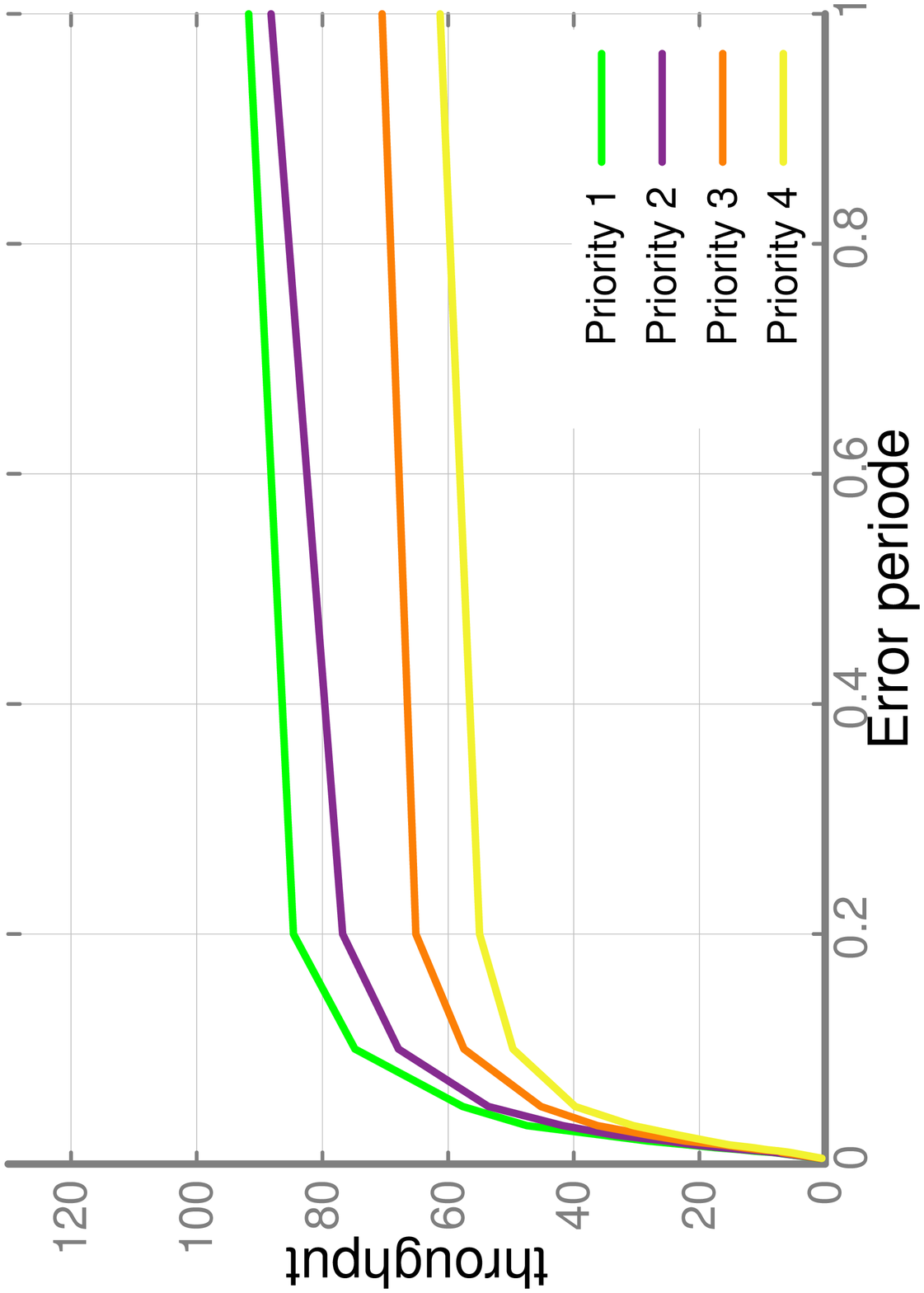}\label{fig:THR_ERR_80211p}}
\hskip -4ex
\subfigure[\small traffic rate $\lambda\!=\! 1000$ 802.11p]{\includegraphics[scale=.15]{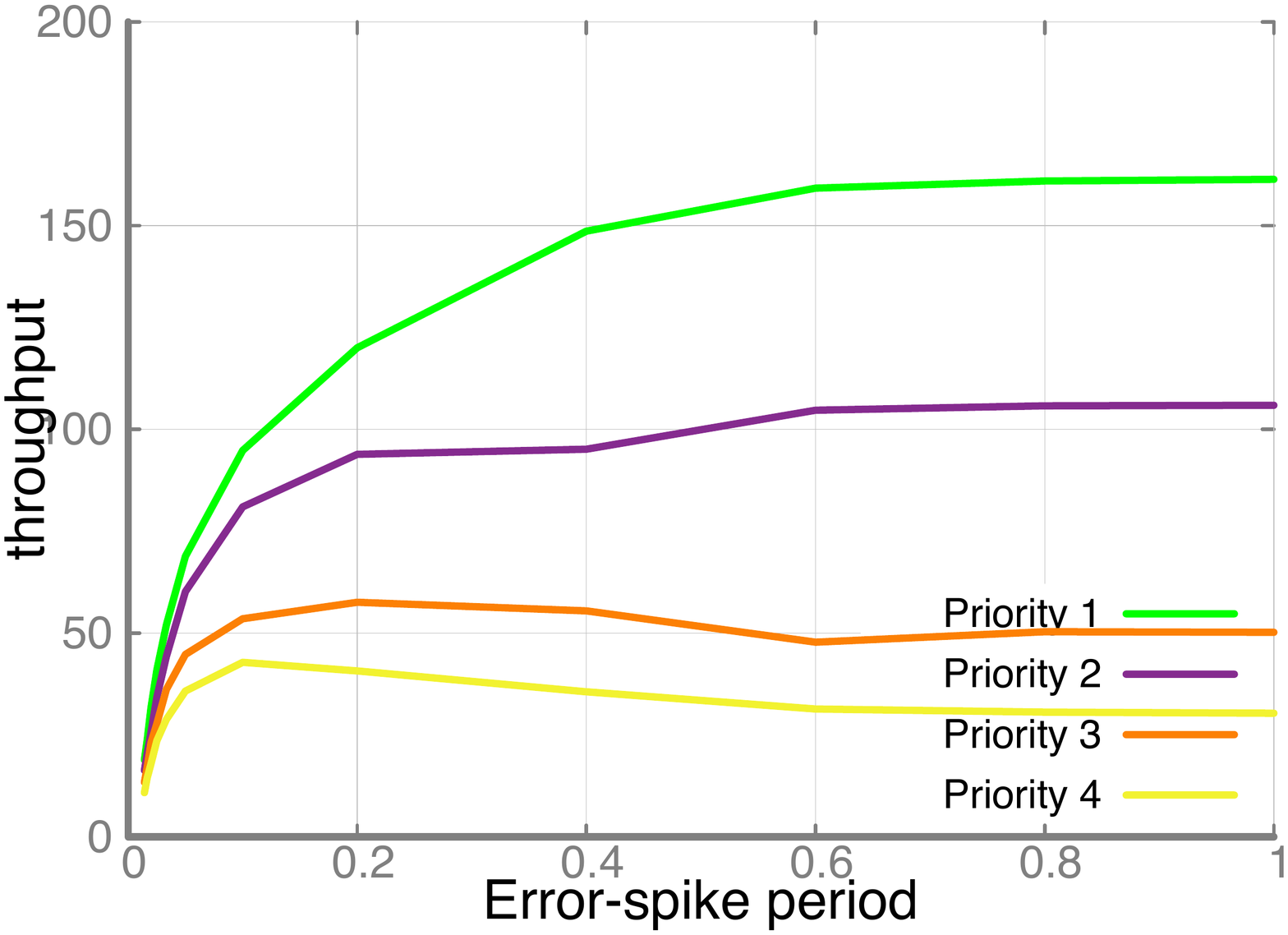}\label{fig:THR_ERR2_80211p}}
\caption{Impact of network dimension and noisy-channel on throughput and BTR for  802.11 and 802.11p networks}
\label{fig:experiments}
\end{center}
\end{figure*}


\subsection{Results}

\paragraph{Impact of network dimension and incoming traffic.} Figure~\ref{fig:THR_80211} and~\ref{fig:BTR_80211}  depict  the estimated  THR and BTR for the 802.11 model computed (for different network dimensions) in function of the traffic arrival rate $\lambda$ (i.e. under a non-saturated regime)   and assuming the channel is not affected by errors (\emph{ideal channel}). Both the THR and  the BTR  exhibit an asymptotic profile (the faster the inter-arrival rate the higher the throughput, resp. the  BTR) and both are  upper bounded by the \emph{nominal maximum throughput under saturated regime} (which is $\sim\! 640$ packets per second), resp. the \emph{saturated nominal maximum BTR}~\footnote{the BTR of a network using 802.11  RTS/CTS MAC has a   \emph{nominal maximum upper bound} ($\mathit{NomMaxBTR}$) corresponding with an ideal (hypothetical) situation where a continuos flow of DATA packets are transmitted without inter-arrival delay and in absence of collisions. In such situation the BTR is given by the ratio  of  busy-time over total-time for successfully  transmitting, where the latter  is given the sum of delays of sequence   \texttt{DIFS}-RTS-\texttt{SIFS}-CTS-\texttt{SIFS}-DATA-\texttt{SIFS}-ACK sequence, that is $\mathit{NomMaxBTR}\! =\! TXtime/ (\mathit{SENSEtime}\!+\!\mathit{TXtime})$, where $\mathit{TXtime}\!=\!RTS\!+\!CTS\!+\!DATA\!+\!ACK$, is the total transmission time and $\mathit{SENSEtime}\!=\!\texttt{DIFS}\!+\!3\texttt{SIFS}$ is the total carrier-sensing time. From parameters in Table~\ref{tab:timeparams}, we have that $\mathit{NomMaxBTR}\!=\!80/88\!=\! 0.909$, meaning that the optimal channel utilisation for a WiFi based on RTS/CTS 802.11 MAC cannot trespass $\sim\!90\%$.}
Figure~\ref{fig:THR_80211p} and~\ref{fig:BTR_80211p},   refer to the same kind of experiment for  the 802.11p model (for a network with $N=2$ nodes). Plots in  Figure~\ref{fig:THR_80211p} highlight the effect of prioritised management of data traffic: the higher the traffic the larger the difference between high priority and low priority throughput. Figure~\ref{fig:BTR_80211p} shows the BTR as well as the ratio between the time the channel is idle (over the total  observation time) and the ratio between the time the channel is garbled (collision).

\emph{Impact of faulty channel.} 
Figure~\ref{fig:THR_ERR_80211}, ~\ref{fig:THR_ERR2_80211},  ~\ref{fig:THR_ERR_80211p}  and~\ref{fig:THR_ERR2_80211p}  report on    assessing the effect that a faulty channel has on the network performances.  The THR and BTR are measured  in  function of the period  of an error spike (determined by the rate of transition $\mathit{EnteringErrorSpike}$). Results show that the throughput increases with the error-spike period    however (for the 802.11 model under saturated regime Figure~\ref{fig:THR_ERR_80211}) the THR  is unaffected by the network  dimension (identical plot for different number of stations) indicating that the performances gradient induced by the network dimension in case of ideal channel (Figure~\ref{fig:THR_80211}) fades away in presence of a faulty channel. Conversely under a  non-saturated regime (802.11 model Figure~\ref{fig:THR_ERR2_80211}) the error-spike period  affects the throughput differently depending on the traffic arrival rates.  Finally Figure~\ref{fig:THR_ERR_80211p}  and~\ref{fig:THR_ERR2_80211p} show the effect of the error-spike on the prioritised throughput in a 802.11p network  showing that the effect of a faulty channel, in terms of the gradient between higher and lower priority throughput is more evident under a high traffic regime (Figure~\ref{fig:THR_ERR2_80211p}) than under a low traffic regime (Figure~\ref{fig:THR_ERR_80211p}).

\section{Conclusion}
\label{sec:conclusion}
We presented a performance modelling study of two versions of MAC protocol for wireless networks: the 802.11 MAC and its prioritised extension 802.11p devoted to VANETS. We developed our models using a high-level stochastic Petri nets formalism which allowed us to encode the complexity of the 802.11 prioritised scheme in a model of reasonable size. The models we presented are highly configurable and allow for the analysis of the performance of networks in different respect (traffic conditions, ideal or imperfect channel, network dimension).   We  analysed the performance on the network models by means of statistical model checking based on the HASL specification language. 
Future work include the extension of the models to  $N$-hops topologies, which would allow us to take into account the effect of routing on the performances of a given network.


\bibliographystyle{plain}
\bibliography{ms}
\newpage
\begin{Appendix}

\section{Parameters of the SSN model of the 802.11 and  802.11p MAC}  

\begin{table}[htp]
\tiny
\begin{center}
\begin{tabular}{|c|C{4.5cm}|c|}

\hline
name & meaning & value ($\times\!10\mu s$)\\
\hline
\texttt{aslot} & the time unit of the backoff procedure & 2\\
\texttt{nStations} & \# of stations & 2\\
\texttt{packetSizeInAslot} & \# of  time slot for sending a DATA packet  & variable\\
\texttt{DIFS} & the length of the DCF interframe space & 5\\
\texttt{SIFS} & the length of a short interframe space & 1\\
\texttt{vuln} & delay to  switch radio from RX-to-TX mode& 2\\
\texttt{RTS} & the time it takes to send a RTS packet & 16\\
\texttt{CTS} & the time it takes to send a CTS packet & 11\\
\texttt{ACK} & the time it takes to send an aknowledgement packet & 11\\
\texttt{timeout} & delay  a station waits for  handshaking packet & 5\\
\texttt{CWmin} &  min. size of the contention window (in aslot) & 15\\
\texttt{backoffMax} &  \# of TX attempts before dropping a packet & 6\\
\texttt{deadline} & \# of time slots for entering the livelock (end) state & 5000\\
\hline
\end{tabular}
\end{center}
\caption{Timing parameters for the SSN models of RTS/CTS}
\label{tab:timeparams}
\end{table}


\begin{table}[htp]
\begin{center}
\begin{tiny}
\begin{tabular}{|c|c|c|}
\hline
\bf {AC} & {\bf CWmin} &  {\bf CWmax} \\
\hline
Background (AC\_BK) & \texttt{aCWmin} & \texttt{aCWmax} \\
Best effort (AC\_BE) & \texttt{aCWmin} & \texttt{aCWmax} \\
Video (AC\_VI) & [(\texttt{aCWmin}+1)/2]-1 & \texttt{aCWmin} \\
Voice (AC\_VO) & [(\texttt{aCWmin}+1)/4]-1 & [(\texttt{aCWmin}+1)/2]-1 \\
\hline
\end{tabular}
\end{tiny}
\end{center}
\caption{Contention windows boundaries for  ECDA access categories.}
\label{tab:contentionwindowpar}
\end{table}%

\begin{table}[htp]
\begin{center}
\begin{tiny}
\begin{tabular}{|c|p{5cm}|c|c|}
\hline
\bf  {\bf name}  &  \centering{\bf description} &  {\bf definition}  & {\bf ordered} \\
\hline
$PT$ & \centering{Packet Type} & $PT\!=\!\{rts,cts,data,ack\}$ & YES \\
$St$ &\centering{ Station ID} & $St\!=\!s\{1, \ldots, N\}$ & NO \\
$Pr$ & \centering{Priority level (of a packet)} & $Pr\!=\!p\{1, \ldots, 4\}$ & NO \\
$TxCount$ & {max. num. of re-transmissions before dropping a packet} & $TxCount\!=\! tx\{1, \ldots, 20\}$ & YES \\
$\mathit{BackoffStage}$ & backoff counter domain   & $\mathit{BackoffStage}\!=\! bs_1\cup \ldots \cup bs_9$ & YES \\
$bs_1$ & backoff counter domain stage 1  & $bs_1\!=\! b\{1,\ldots, 4\}$ & YES \\
$bs_2$ & backoff counter domain stage 2  & $bs_2\!=\! b\{5,\ldots, 8\}$ & YES \\
$bs_3$ & backoff counter domain stage 3  & $bs_3\!=\! b\{9,\ldots, 16\}$ & YES \\
$bs_4$ & backoff counter domain stage 4  & $bs_4\!=\! b\{17,\ldots, 32\}$ & YES \\
$bs_5$ & backoff counter domain stage 5  & $bs_5\!=\! b\{33,\ldots, 64\}$ & YES \\
$bs_6$ & backoff counter domain stage 6  & $bs_6\!=\! b\{65,\ldots, 128\}$ & YES \\
$bs_7$ & backoff counter domain stage 7  & $bs_7\!=\! b\{129,\ldots, 256\}$ & YES \\
$bs_8$ & backoff counter domain stage 8  & $bs_8\!=\! b\{257,\ldots, 512\}$ & YES \\
$bs_9$ & backoff counter domain stage 9  & $bs_9\!=\! b\{513,\ldots, 1024\}$ & YES \\
\hline
\end{tabular}
\end{tiny}
\end{center}
\caption{Color classes for the SSN model of the 802.11p protocol.}
\label{tab:colorclasses}
\end{table}%

\begin{table}[htp]
\begin{center}
\begin{tiny}
\begin{tabular}{|c|c|c|}
\hline
\bf  {\bf name}  &  \centering{\bf description} &  {\bf definition}    \\
\hline
$SR$ & \centering{Sender-Receiver} & $SR\!=\!St\times St$ \\
$SP$ &\centering{ Station-Priority} & $SP\!=\!St\times Pr$   \\
$BP$ &\centering{ Backoff-Priority} & $BP\!=\!St\times St\times Pr$   \\
$P$ &\centering{Packet sending} & $P\!=\!St\times St\times Pr\times PT$   \\
$SRB$ &\centering{Sender-Receiver-BackoffStage} & $SRB\!=\!St\times St\times \mathit{BackoffStage}$  \\
$BPMap$ &\centering{Mapping-Priority-Backoff} & $BPMap\!=\!Pr\times TxCount\times \mathit{BackoffStage}$  \\
$SRPcount$ &\centering{Sender-Recevier-Priority-Count} & $SRPcount\!=\!St\times St\times Pr\times  \mathit{TxCount}$  \\
\hline
\end{tabular}
\end{tiny}
\end{center}
\caption{Color domains for the SSN model of the 802.11p protocol.}
\label{tab:colordomain}
\end{table}%

\begin{table}[htp]
\begin{center}
\begin{tiny}
\begin{tabular}{|m{0.75cm}|L{7.25cm}|}
\hline
{\bf marking}  &   {\bf definition} \\ 
\hline
$\mathit{BM_0}$ & <pr1,tx1,bs1>+<pr1,tx2,bs1+bs2>+<pr1,tx3,bs1+bs2>+<pr1,tx4,bs1+bs2>  
 +<pr2,tx1,bs1>+<pr2,tx2,bs1+bs2>+<pr2,tx3,bs1+bs2+bs3> + 
 +<pr2,tx4,bs1+bs2+bs3>+<pr2,tx5,bs1+bs2+bs3>+<pr3,tx1,bs1+bs2> 
 + <pr3,tx2,bs1+bs2+bs3>+<pr3,tx3,bs1+bs2+bs3+bs4>   
 +<pr3,tx4,bs1+bs2+bs3+bs4+bs5> + <pr3,tx5,bs1+bs2+bs3+bs4+bs5+bs6> 
 +<pr3,tx6,bs1+bs2+bs3+bs4+bs5+bs6+bs7>   
+ <pr3,tx7,bs1+bs2+bs3+bs4+bs5+bs6+bs7+bs8> 
 +<pr3,tx8,bs1+bs2+bs3+bs4+bs5+bs6+bs7+bs8+bs9>
 + <pr3,tx9,bs1+bs2+bs3+bs4+bs5+bs6+bs7+bs8+bs9>  
+ <pr3,tx10,bs1+bs2+bs3+bs4+bs5+bs6+bs7+bs8+bs9>  
 +<pr4,tx1,bs1+bs2+bs3>+<pr4,tx2,bs1+bs2+bs3+bs4> 
 +<pr4,tx3,bs1+bs2+bs3+bs4+bs5>+<pr4,tx4,bs1+bs2+bs3+bs4+bs5+bs6> 
 +<pr4,tx5,bs1+bs2+bs3+bs4+bs5+bs6+bs7> 
 +<pr4,tx6,bs1+bs2+bs3+bs4+bs5+bs6+bs7+bs8>  
+<pr4,tx7,bs1+bs2+bs3+bs4+bs5+bs6+bs7+bs8+bs9>
+ <pr4,tx8,bs1+bs2+bs3+bs4+bs5+bs6+bs7+bs8+bs9>  
+ <pr4,tx9,bs1+bs2+bs3+bs4+bs5+bs6+bs7+bs8+bs9> \\
\hline
\end{tabular}
\end{tiny}
\end{center}
\caption{Invariant marking of place $\mathit{BackoffMapping}$.}
\label{tab:marking}
\end{table}%
  
\end{Appendix}

\end{document}